\documentclass[5p, sort&compress]{elsarticle} 
\usepackage{graphicx}
\usepackage{amssymb}
\usepackage{epstopdf}
\usepackage[amssymb,thinqspace,thinspace]{SIunits}
\usepackage{amsmath}
\usepackage{booktabs}
\usepackage{rotating}
\usepackage{multirow}
\usepackage{psfrag}
\usepackage{adjustbox}
\usepackage{lineno}
\usepackage{csquotes}

\usepackage{color}
\journal{Materials Science and Engineering: A}

\begin{document}
\begin{frontmatter}
\title{Design of a High Strength, High Ductility 12 wt\% Mn Medium Manganese Steel With Hierarchical Deformation Behaviour}
\author[IC]{T. W. J. Kwok}
\author[IC]{K. M. Rahman}
\author[IC]{X. Xu}
\author[IC]{I. Bantounas}
\author[ISIS]{J. F. Kelleher}
\author[UNT]{S. Dasari}
\author[UNT]{T. Alam}
\author[UNT]{R. Banerjee}
\author[IC]{D. Dye}

\address[IC]{Department of Materials, Royal School of Mines, Imperial College London, Prince Consort Road, London, SW7 2BP, United Kingdom}
\address[ISIS]{ISIS Neutron and Muon Source, Rutherford Appleton Laboratory, Didcot, OX11 0QX, United Kingdom}
\address[UNT]{Department of Materials Science and Engineering, University of North Texas, 1155 Union Circle \#305310, Denton, TX 75203-5017} 

\begin{abstract}

A novel medium Mn steel of composition Fe-12Mn-4.8Al-2Si-0.32C-0.3V was manufactured with 1.09 GPa yield strength, 1.26 GPa tensile strength and 54\% elongation. The thermomechanical process route was designed to be industrially translatable and consists of hot and then warm rolling followed by a 30 min intercritical anneal. The resulting microstructure comprised of coarse elongated austenite grains in the rolling direction surrounded by necklace layers of fine austenite and ferrite grains. The tensile behaviour was investigated by in-situ neutron diffraction and the evolution of microstructure studied with Electron Backscattered Diffraction (EBSD). It was found that the coarse austenite grains contributed to the first stage of strain hardening by transforming into martensite and the fine austenite necklace grains contributed to the second stage of strain hardening by a mixture of twinning and transformation induced plasticity (TWIP and TRIP) mechanisms. This hierarchical deformation behaviour contributed to the exceptional ductility of this steel.
\end{abstract}

\end{frontmatter}


\section{Introduction}
\label{Intro}

Mn-containing steels exploiting twinning to provide hardening and therefore ductility date back to Hadfield steels, discovered by Sir Robert Hadfield in 1888 \cite{Hadfield1888}. Current high Mn steels or TWIP steels are fully austenitic steels that deform by forming nano-scale twins in addition to dislocation glide \cite{DeCooman2011,DeCooman2018}. The deformation twins are thought to provide a dynamic Hall-Petch effect where the twins continuously subdivide the grain \cite{Bouaziz2008} and provide high strain hardening rates of up to 3 $\usk\giga\pascal$ and exceptional ductility, often in the excess of 50\% \cite{Rahman2015}. 


These mechanical properties made TWIP steels very attractive for energy absorbing applications such as automotive crash assemblies and armour plate \cite{Rahman2014}. However, the high Mn content, up to 30 wt\% Mn \cite{Chen2017}, poses many problems in secondary steel melting and downstream processing such as (1) macrosegregation in slabs resulting in poor hot roll-ability \cite{Bausch2013}, (2) hydrogen cracking \cite{DeCooman2018} and (3) poor weldability, especially in Zn-coated sheets \cite{DeCooman2018, Chen2017, Papadimitriou2017}. Medium Mn steels containing $6-12$ wt\% Mn have been recently gaining significant attention in an attempt to solve this problem. The alloying concept has therefore been to significantly reduce the amount of Mn while retaining the exceptional mechanical properties of high Mn steel.

Many researchers have been succesful in retaining the TWIP effect with significantly lowered Mn content \cite{Kim2019, Lee2015e, Lee2014, Hu2017b}. Lee \textit{et. al.} \cite{Lee2015e} reported twins in a medium Mn steel with only 6 wt\% Mn. However, the unique feature of medium Mn steels is not only the ability to retain the TWIP effect with less alloyed Mn but the ability to exhibit both TWIP and TRIP effects \cite{Lee2014, Lee2015e, Lee2016}. TWIP and TRIP effects have been observed to occur either in succession (\textit{i.e.} TWIP then TRIP), or simultaneously. Sohn \textit{et. al.} \cite{Sohn2014a} showed that twinning and martensitic transformation were able to occur simultaneously within the same austenite grain. However, simultaneous TWIP and TRIP is not as commonly reported as the successive TWIP then TRIP mechanism. In the successive TWIP then TRIP mechanism, the steel initially deforms by the TWIP mechanism and with increasing strain, twin intersections serve as potent nucleation sites for strain induced martensitic transformation and the TRIP mechanism is activated \cite{Lee2014}.

Whilst the successive TWIP+TRIP effect has been studied in metastable austenitic stainless steels \cite{Tian2017}, the conditions in which this effect is active in medium Mn steel are less well understood. It is gathered that the TWIP+TRIP effect depends strongly on both alloy chemistry and microstructure \cite{Lee2014,Lee2015d,Sun2018,Kim2019}, which is unsurprising since these two factors significantly influence stacking fault energy (SFE) and austenite stability which govern the TWIP and TRIP effects respectively. In turn, thermomechanical processing influences the microstructure and phase chemistry through element partitioning during processing and heat treatment in the intercritical $\gamma + \alpha$ temperature region.

Most thermomechanical processing methods for medium Mn steel in the literature aim to overcome the problem of slow Mn diffusivity in both austenite and ferrite \cite{Lee2011a,Nakada2014,Lee2013c}. Incomplete partitioning of Mn may lead to undesired SFE, austenite stability, non-equilibrium phase fractions and ultimately poor mechanical properties. To overcome this problem, one of the most common processing approaches are of hot roll - quench - cold roll - intercritical anneal (IA) type \cite{Lee2014,Lee2015b,Sohn2014a}. Cold rolling prior to IA is beneficial as it introduces a large density of defects which allows for recrystallisation and rapid partitioning of elements within a short period of time during intercritical annealing. Lee \textit{et. al.} \cite{Lee2011a} showed that in a 6 wt\% Mn alloy, partitioning of Mn was nearly complete, \textit{i.e.} reaching thermodynamic equilibrium, after cold rolling and IA for only 180 s. Without cold rolling, the desired extent of partitioning may only be reached after several hours of annealing or not at all \cite{Lee2015c,Nakada2014}. Warm rolling as an alternative to cold rolling of medium Mn steel has been explored \cite{Hu2017b,Hu2019}. Hu \textit{et. al.} \cite{Hu2017b} developed a 7 wt\% Mn steel with 63\% ductility through warm rolling at 700 \degree C and a 5 hr IA also at 700 \degree C. The Mn content in austenite grains was found to be between 9 and 10.9 wt\%, which is similar to the prediction by Thermo-Calc (11 wt\%) under equilibrium conditions, demonstrating the potential of warm rolling to replace cold rolling.

However, continuous thermomechanical processing without the need for quenching or cold rolling has not received much attention although it is the most economical and widespread method of producing sheet steel in industry. In this study on a 12 wt\% Mn medium Mn steel, we present a novel continuous thermomechanical processing route and examine the evolution of microstructure with each processing step. The tensile behaviour and deformed microstructures were also examined with in-situ neutron diffraction and EBSD.

\section{Experimental Procedures}

\begin{figure*}[h]
	\centering
	\includegraphics[width=\linewidth]{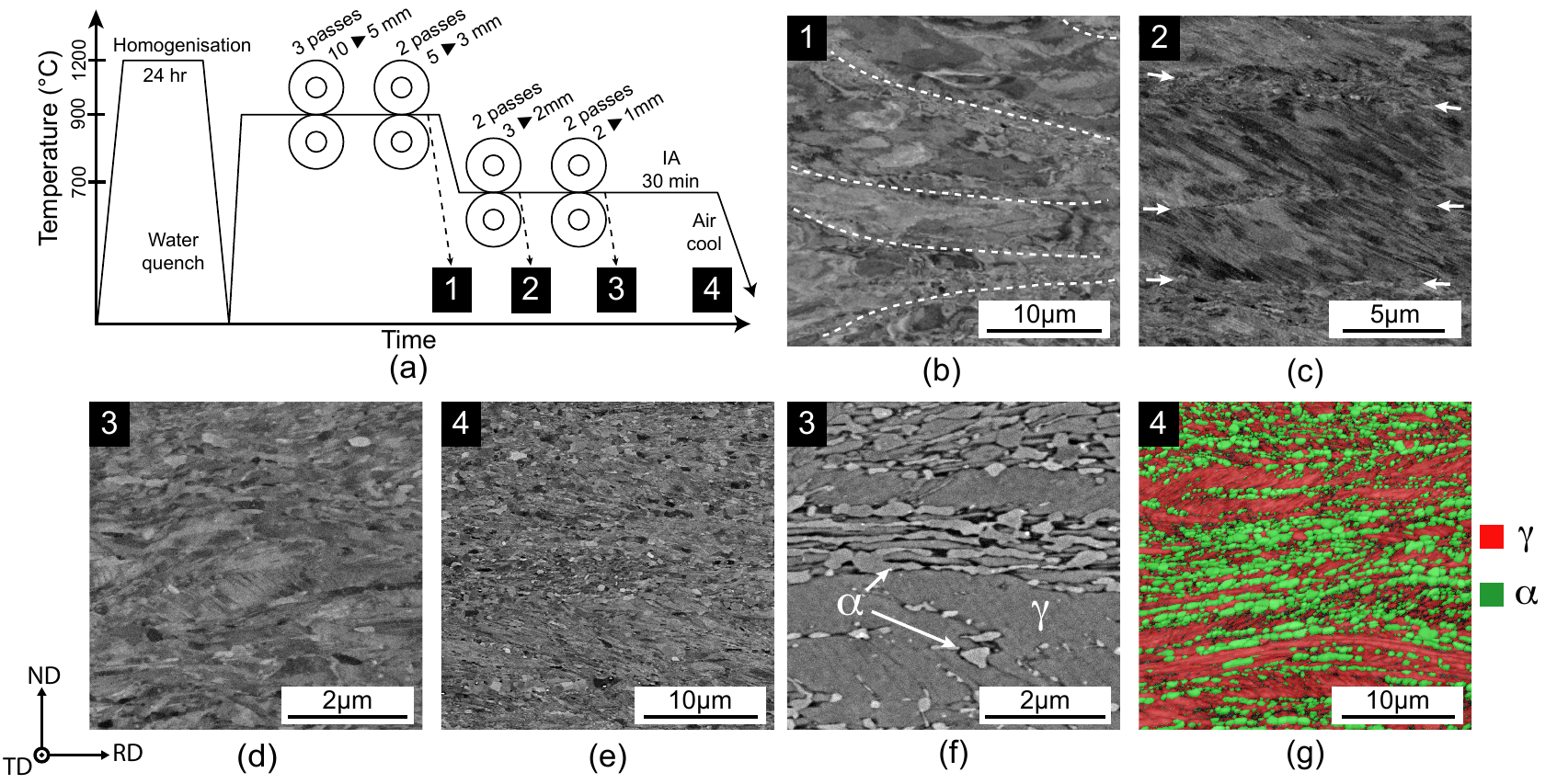}
	\caption{(a) Thermomechanical processing schedule showing extracted samples, \textit{i.e.} stages 1-4. Backscattered electron (BSE) micrographs of the steel after (b) hot rolling at 900 \degree C where dotted lines indicate elongated prior austenite grain boundaries, (c) 2 passes at 700 \degree C where white arrows point to the first necklace layers which have started to form, (d) 4 passes at 700 \degree C and (e) IA at 700 \degree C for 30min. (f) Etched secondary electron micrograph of (d) and (g) EBSD IQ+PM scan of (e).}
	\label{fig:processing-micrographs}
\end{figure*}

The steel used in this work was of nominal composition Fe-12Mn-4.8Al-2Si-0.3C-0.3V (wt\%) prepared using pure elements and vacuum arc melted to produce 60$\times$10$\times$10 mm bars. The bars were quartz encapsulated and homogenised in low pressure Ar at 1200 \degree C for 24 h and water quenched.  Thermomechanical processing was conducted at 900 \degree C in five passes (hot rolling), 700 \degree C in four passes (warm rolling) and annealed at 700 \degree C for 30 min before air cooling to room temperature. All reductions were approximately 20\% per pass and the final thickness of the strip was approximately 1 mm. To chart the evolution of microstructure along the processing schedule, several bars were quenched after the 5th, 7th and 9th pass before IA.  A schematic of the rolling schedule and specimens used for microstructual analysis are shown in Figure \ref{fig:processing-micrographs}a.

Sub-sized tensile specimens with gauge dimensions of 19$\times$1.5$\times$1 mm were cut \textit{via} Electric Discharge Machining (EDM) such that the gauge was parallel to the rolling direction. Tensile testing was conducted on an Instron 5960 load frame equipped with a 30 kN load cell at a nominal strain rate of $10^{-3} \: s^{-1}$. To investigate the microstructural changes during uniaxial deformation, interrupted tensile tests were conducted to 0.05, 0.18 and 0.28 true strain ($\epsilon_t$).

Scanning Electron Microscopy (SEM), EBSD and Energy Dispersive Spectroscopy (EDS) measurements were performed on a Zeiss Sigma FE-SEM equipped with a Bruker EBSD detector and Bruker XFlash 6160 EDS detector. Transmission Electron Microscopy (TEM) was conducted on a JEOL JEM-2100F. TEM samples were obtained by Focused Ion Beam (FIB) milling and lifting out a foil in a FEI Helios Nanolab 600 focused ion beam workstation. Specimens for Atom Probe Tomography (APT) were prepared by FIB milling and lifting out a foil before mounting the small sections of the samples on suitable holders for analysis. 

The APT experiments were conducted on a CAMECA local electrode atom probe 5000XS instrument. All experiments were performed at a temperature of 30 K with a target evaporation of 0.6\% and laser power of 30 nJ with a steady-state applied DC voltage. APT data reconstruction and analysis were carried out using CAMECA IVAS 3.8.4 software. The mass spectrum for the APT reconstruction was calibrated based on the average phase compositions obtained from SEM-EDS. The mass-to-charge ratio peaks were assigned as follows: C$^{2+}$ (6, 6.5 Da), C$^+$ (12, 13 Da), CC$^+$ (24, 26 Da), CCC$^+$ (18, 18.5, 36 Da), Al$^{3+}$ (9.0 Da), Al$^{2+}$ (13.5 Da), Al$^+$ (27 Da), Si$^{2+}$ (14, 14.5, 15 Da), V$^{3+}$ (17 Da), V$^{2+}$ (25.5 Da), Mn$^{2+}$ (27.5 Da), Mn$^+$ (55.0 Da), Fe$^{2+}$ (27, 28, 28.5, 29 Da), Fe$^+$ (54, 56, 57, 58 Da) and CCFe$^{2+}$ (40 Da). The overlap in Fe$^{2+}$ peaks with Si$^+$ was ignored as the intensities matched with the relative abundance. The 27 Da peak was proportionally split between Al$^+$ and Fe$^{2+}$ based on the relative intensities of remaining three Fe$^{2+}$ peaks. While it has been shown that lower specimen temperature is preferred for better quantitative analysis \cite{Yamaguchi2009}, Takahashi \textit{et. al.} \cite{Takahashi2011} and Miyamoto \textit{et. al.} \cite{Miyamoto2012} have observed an increase in carbon concentration in steels with decrease in specimen temperature. In this study, a relatively lower specimen temperature (30 K) was used. With these considerations in mind, it is acknowledged that there will be some uncertainty in the elemental compositions during the APT reconstruction of the specimen. Nevertheless, APT provides superior spatial resolution and the results are still useful in the investigation of relative element partitioning between the different phases in medium Mn steel, which is the principal focus.

\begin{figure}[b!]
	\centering
	\includegraphics[width=\linewidth]{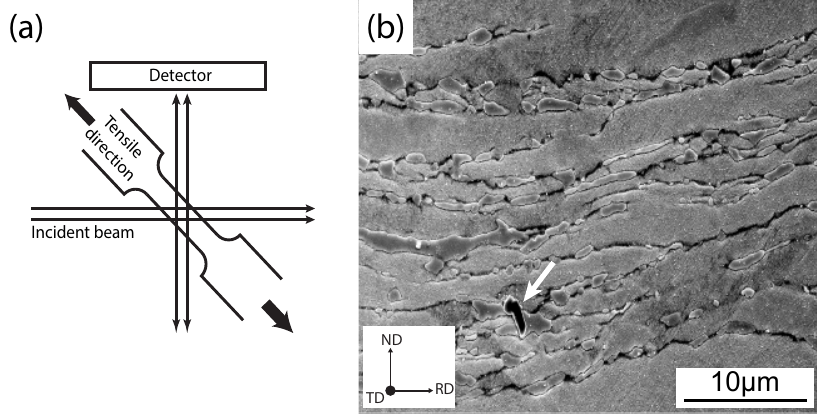}
	\caption{(a) Schematic of the in-situ neutron diffraction tensile experiment. (b) Secondary electron micrograph of the neutron tensile specimen at 0\% strain. White arrow indicates a large precipitate.}
	\label{fig:neutron-set-up-schematic}
\end{figure}

In order to further study the tensile behaviour of the steel, another ingot of the same composition measuring 60$\times$23$\times$23 mm was homogenised in a vacuum furnace at 1200 \degree C for 24 h, furnace cooled and processed according to a similar rolling schedule. The ingot was hot rolled at 900 \degree C from 23 mm to 12.5 mm in 4 passes of approximately $15-20$\% reduction per pass and warm rolled at 700 \degree C from 12.5 mm to 6 mm in 4 passes of approximately $15-20$\% reduction per pass before a 30 min IA at 700 \degree C.

A tensile specimen for neutron diffraction with gauge dimensions of 50$\times$6$\times$6 mm and sub-sized tensile specimens were similarly cut \textit{via} EDM with the gauge parallel to the rolling direction from the same strip. The in-situ neutron diffraction experiment was conducted at the ENGIN-X diffractometer at the ISIS Neutron and Muon Source,  Rutherford Appleton Laboratory (Oxfordshire, UK). A tensile test was performed at room temperature using a 100 kN load frame with the loading axis placed on the bisection of the incident and diffracted beams (Figure \ref{fig:neutron-set-up-schematic}). The tensile test was performed in load control in the elastic region and displacement control mode during plastic elongation. Loading was paused for each measurement.  Diffracted peaks obtained in the detector correspond to the loading of grains with scattering plane normal to the loading direction and were simultaneously collected by the north and south detector banks, both positioned at the horizontal scattering angles of $\pm \: 90\degree$. In this study, only the patterns recorded by the north detector bank were used, which depicts the longitudinal lattice strains of the lattice planes with the normal in the tensile direction. The elastic lattice strain, $\epsilon^{hkl}$, of grains in scattering condition \{hkl\} can be obtained from shift in peak positions according to $\epsilon^{hkl} = (d^{hkl}-d_0^{hkl})/{d_0^{hkl}}$, where $d_0^{hkl}$ and $d^{hkl}$ are the unstrained and strained lattice parameters respectively.

\section{Results}

\subsection{Microstructural evolution during processing}

Medium Mn steels have great flexibility in microstructural design and can exhibit very different microstructures depending on the processing route. At 900 \degree C, the steel was fully austenitic and at stage 1 in Figure \ref{fig:processing-micrographs}b, the microstructure consisted of elongated grains, suggesting that recrystallisation of austenite was limited at this temperature. When the temperature was reduced to 700 \degree C and after the first 2 warm passes (stage 2), a single layer of very fine grains was observed to have nucleated on the Prior Austenite Grain Boundaries (PAGB). After stage 3, it was observed that additional layers of fine grains had formed along the PAGB, partially consuming the prior austenite grain. Etching in 2\% nital solution clearly revealed that the layers consisted of two phases (Figure \ref{fig:processing-micrographs}f) which were likely austenite and ferrite, with ferrite also nucleating along shear bands in the deformed parent austenite grain. The final microstructure after the full processing schedule could not easily be discerned in the BSE micrograph but the EBSD image quality and phase map (IQ+PM) clearly showed bands of elongated austenite and also bands of fine austenite and ferrite grains. The grains in the fine grained region will be termed necklace grains as they formed in layers or necklaces, closely resembling the necklace-type recrystallisation mechanism \cite{Ponge1998}. The retained elongated austenite grains will be termed the core austenite grains because the necklace recrystallisation mechanism was incomplete, leaving an austenite \enquote{core} in the middle of a prior austenite grain. 
 
\begin{table}
	\caption{Composition of the bulk steel measured by ICP and individual phases measured by SEM-EDS and APT. Uncertainties shown in parentheses. $\dagger$The C content of the bulk steel was measured by Time of Flight Mass Spectrometry (TOFMS). The C content in necklace and core austenite EDS measurements was assumed to be the equal and calculated by the lever rule on the basis that C has minimal solubility in ferrite.}
\begin{adjustbox}{width=\columnwidth,center}\begingroup
	\setlength{\tabcolsep}{3pt} 
    \begin{tabular}{lccccccc}
    	\toprule
         & \multicolumn{5}{c}{Mass percent (\%)}         & \multicolumn{1}{c}{SFE} & \multicolumn{1}{c}{M$_d$ } \\
         & Mn    & Al    & Si    & V     & C$\dagger$     & mJ m$^{-2}$ & \degree C \\
          \midrule
    Bulk (ICP) & 11.99 & 4.84  & 1.98  & 0.28  & 0.32  & -   & - \\
    Bulk (EDS) & 10.3 (0.1) & 4.9 (0.1) & 1.3 (0.1) & 0.2 & 0.32 & - & - \\
        $\gamma$ Thermo-Calc & 13.90 & 4.16 & 1.40 & 0.003 & 0.57 & 36.6 & - \\
    $\gamma$ core (EDS) & 10.0 (0.5) & 4.8 (0.3) & 1.3 (0.1) & 0.2 (0.1) & 0.52 & 36.0  & 250 \\
    $\gamma$ necklace (EDS)& 11 (2) & 4.0 (0.7) & 1.0 (0.4) & 0.3 (0.3) & 0.52  & 31.1 & 225 \\
    $\alpha$ necklace (EDS) & 7.8 (0.8) & 4.6 (0.4) & 1.3 (0.2) & 0.3 (0.3) & 0     & -    & - \\
    \midrule
    & \multicolumn{5}{c}{Atomic percent (\%)}         & \multicolumn{1}{c}{SFE} & \multicolumn{1}{c}{M$_d$ } \\
    & Mn    & Al    & Si    & V     & C$\dagger$     & mJ m$^{-2}$ & \degree C \\
    \midrule
     $\gamma$ Thermo-Calc & 13.07 & 7.96 & 2.57 & 0.003 & 2.45 & 36.6 & - \\
     $\gamma$ necklace (EDS) & 11(1) & 8 (1) & 1.9 (0.7) & 0.3 (0.3) & 2.25  & 31.1 & 225 \\
      $\gamma_1$ necklace (APT) & 13.28 (0.07) & 10.20 (0.06) & 2.77 (0.03) & 0.05 & 1.63 (0.02)  & 36.3 & 273 \\
     $\gamma_2$ necklace (APT) & 13.31 (0.07) & 9.48 (0.06) & 2.97 (0.03) & 0.04  & 1.37 (0.02)  & 30.1 & 298 \\
    $\alpha$ necklace (EDS) & 7.5 (0.8) & 9.0 (0.4) & 2.4 (0.2) & 0.3 (0.3) & 0     & -    & - \\
    $\alpha$ necklace (APT) & 8.28 (0.06) & 14.53 (0.08) & 3.78 (0.04) & 0.03  & 0     & -    & - \\
    \bottomrule
    \end{tabular}\endgroup
\end{adjustbox}
  \label{tab:SEM-EDS-ICP-compositions}%
\end{table}%

\begin{figure}[t!]
	\centering
	\includegraphics[width=\linewidth]{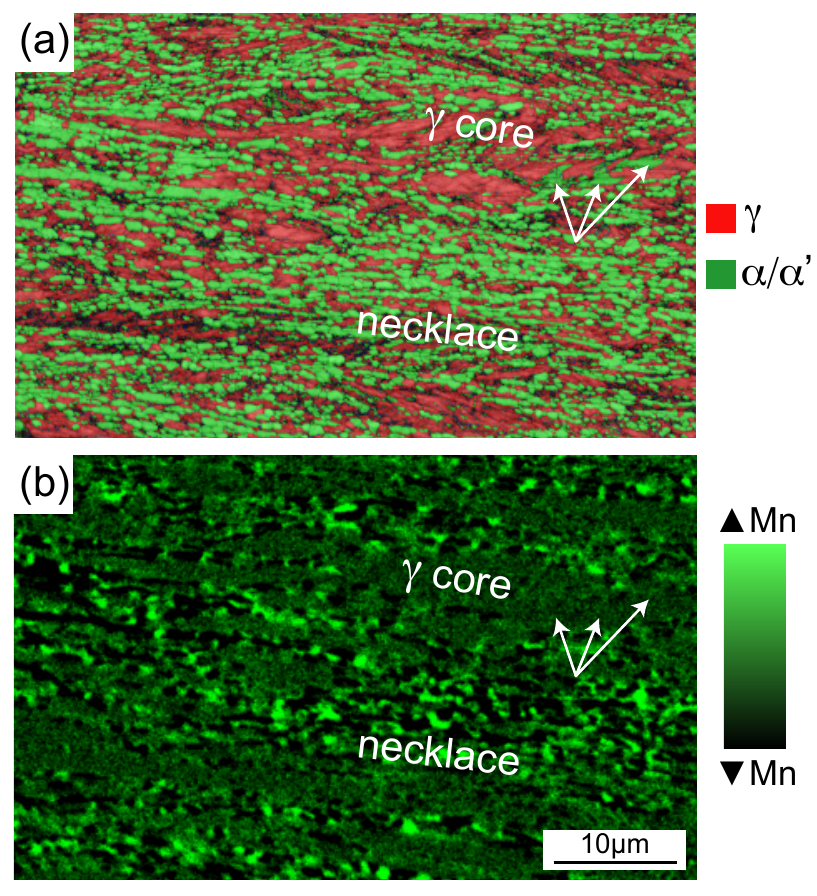}
	\caption{(a) EBSD IQ+PM and corresponding (b) qualitative SEM-EDS map of a specimen processed according to Figure \ref{fig:processing-micrographs}a strained to $\epsilon_t = 0.05$. Locations of strain induced martensite are indicated by the white arrows.}
	\label{fig:sem-eds-mn-map}
\end{figure}

As the deformation mechanisms active in medium Mn steels are heavily dependent on composition, the composition of the studied steel was measured with three different methods at different length scales and is shown in Table \ref{tab:SEM-EDS-ICP-compositions}. The bulk composition was measured by Inductively Coupled Plasma Spectroscopy (ICP) and SEM-EDS. SEM-EDS was also used to map the distribution of Mn in the microstructure (Figure \ref{fig:sem-eds-mn-map}) and to probe the composition of individual core austenite, necklace ferrite and austenite grains. Finally, APT was used to probe the local composition within one necklace ferrite and two necklace austenite grains.

Compositions measured by SEM-EDS were averaged over a minimum of 30 grains. Due to the difficulty in measuring C content in SEM-EDS, the C content in ferrite was assumed to be negligible and C content in austenite was determined by the lever rule. C content in a necklace ferrite grain was indeed confirmed to be negligible through APT as shown in Table \ref{tab:SEM-EDS-ICP-compositions}. The C content in both core and necklace austenite grains were assumed to be equal due to the fast diffusion kinetics of C during partitioning \cite{Kamoutsi2015} but some slight differences between necklace austenite grains was observed in the APT measurement. The SFE and M$_d$ temperature were calculated according to the method developed by Pierce \textit{et. al.} \cite{Pierce2014} and Nohara \textit{et. al.} \cite{Nohara1977,Talonen2005,Sun2018} accordingly. The thermodynamic equilibrium composition of austenite at 700 \degree C was determined through Thermo-Calc using the TCFE 7.0 database based on the bulk composition obtained by SEM-EDS.

\begin{figure}[t!]
	\centering
	\includegraphics[width=\linewidth]{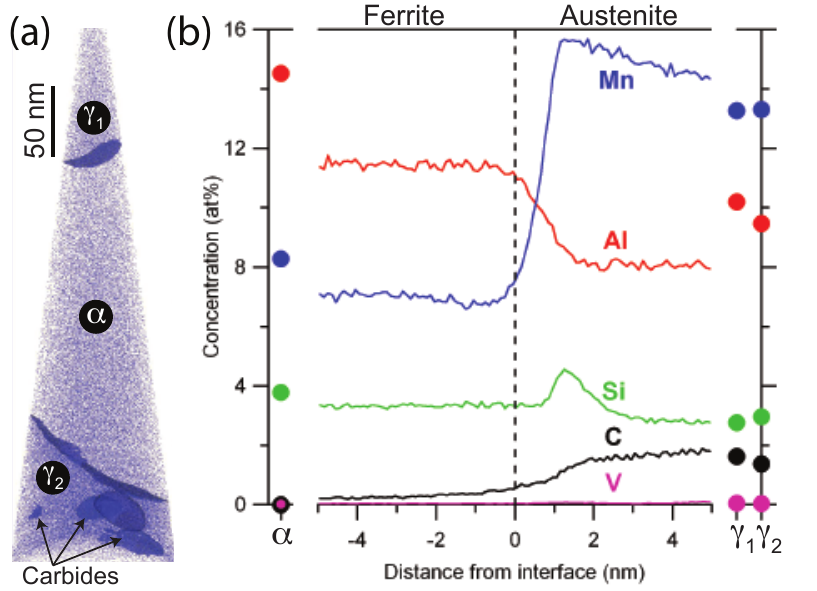}
	\caption{(a) Mn ion map with 9.3 at\% iso-surface obtained from the necklace region. (b) Proxigram analysis of both the top and bottom interfaces. Far field analyses of the compositions of the $\alpha$, $\gamma_1$ and $\gamma_2$ necklace austenite grains are given at the edges of (b).}
	\label{fig:apt-reconstruction-1}
\end{figure}

\begin{figure*}[t!]
	\centering
	\includegraphics[width=\linewidth]{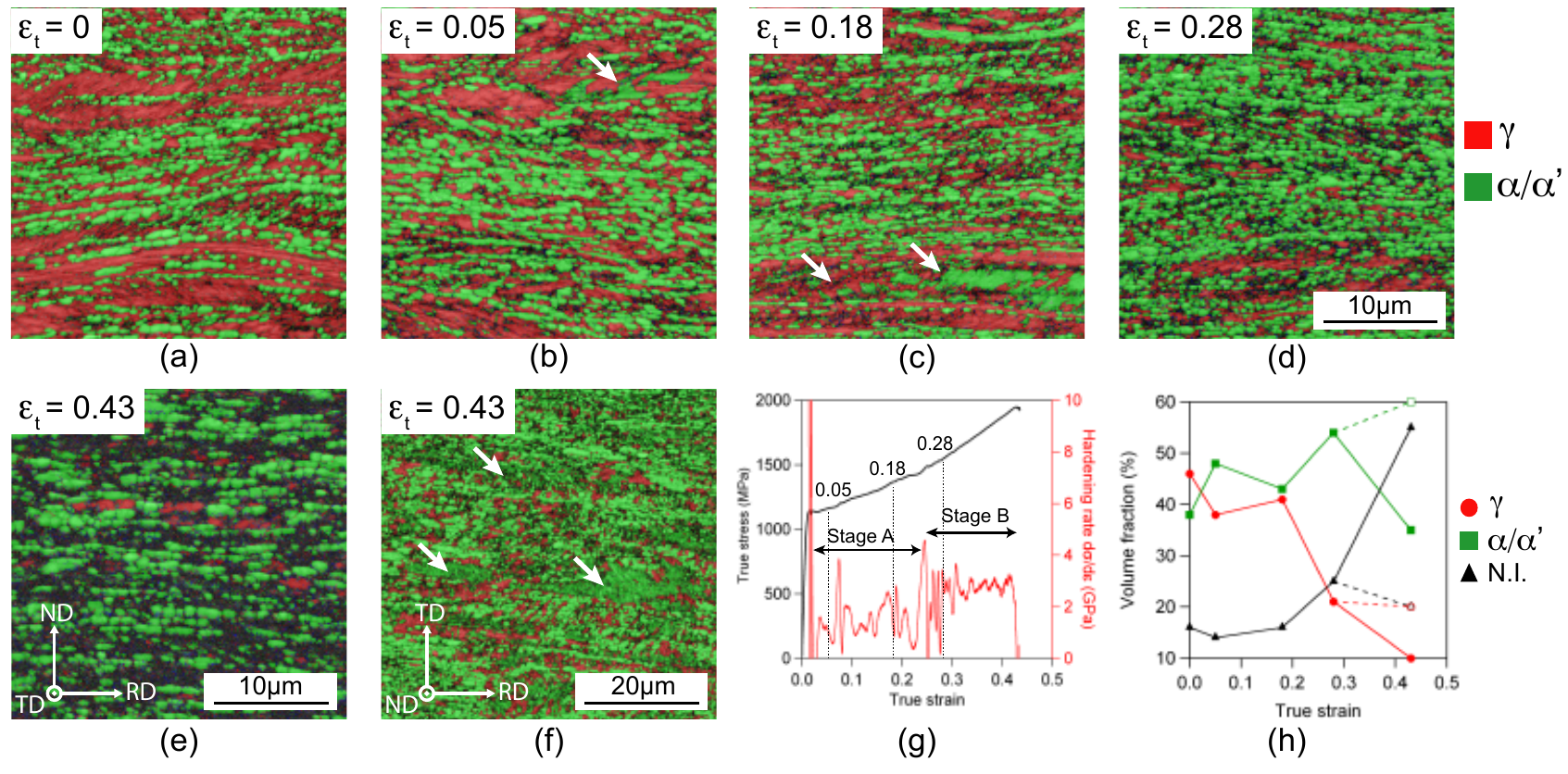}
	\caption{EBSD IQ+PM of interrupted tensile specimens obtained parallel to the transverse direction at (a) $\epsilon_t = 0$, (b) $\epsilon_t = 0.05$, (c) $\epsilon_t = 0.18$, (d) $\epsilon_t = 0.28$ and (e) $\epsilon_t = 0.43$. White arrows indicate strain induced blocky $\alpha$' martensite. (f) EBSD IQ+PM map at $\epsilon_t = 0.43$ but obtained parallel to the normal direction. White arrows indicate locations of completely transformed core austenite grains. (g) True stress-strain curve and hardening rate. (h) Change in austenite, ferrite/martensite and non-indexed (N.I.) area fraction obtained from (a-e), the dotted line and open symbol represents the volume fractions obtained in (f).}
	\label{fig:ebsd-vol-frac-tensile2}
\end{figure*}

\begin{figure}[t!]
	\centering
	\includegraphics[width=.8\linewidth]{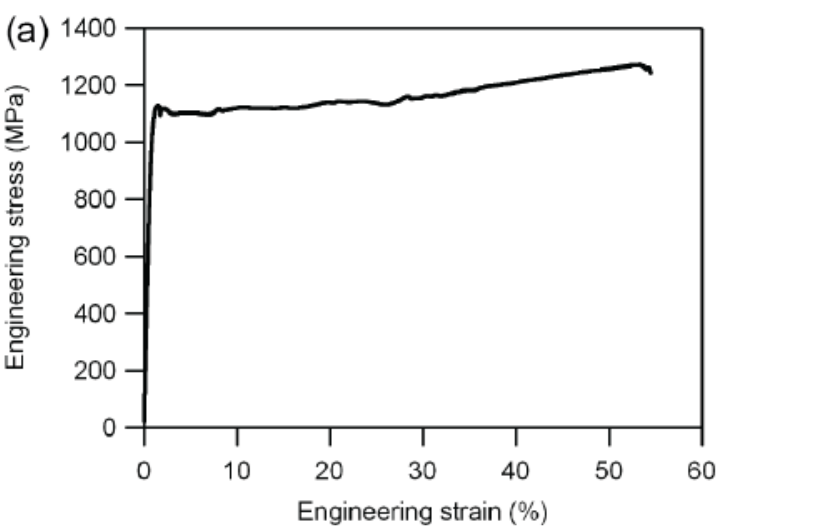}
	\caption{Engineering stress strain curve for the steel studied.}
	\label{fig:tensiles2}
\end{figure}

When comparing the EBSD IQ+PM map and Mn EDS map in Figure \ref{fig:sem-eds-mn-map}, it can be seen that the fine necklace austenite grains were enriched in Mn and the ferrite grains were Mn depleted. The core austenite can be distinguished as having an intermediate level of Mn and from Table \ref{tab:SEM-EDS-ICP-compositions}, the core austenite grains had a very similar composition to the bulk composition. This suggests that partitioning did not occur significantly in these grains. SEM-EDS measurements of the necklace austenite grains showed that they were slightly enriched in Mn and depleted in Al and Si, suggesting that partitioning occurred between the necklace austenite and ferrite grains to a greater degree. However, the composition of the necklace austenite grains measured by APT revealed that the extent of partitioning was much greater than that measured by SEM-EDS. Due to the small size of the austenite and ferrite grains within the necklace region, the electron interaction area associated with SEM-EDS may be larger than a single grain and the composition of any carbides and neighbouring ferrite grains could have been sampled during the SEM-EDS measurement of individual necklace austenite grains. The SEM-EDS measurement of necklace austenite grains can therefore be interpreted as an \enquote{average} composition of these grains. APT also revealed that very little V remained in solution in both phases and likely precipitated out as carbides as shown in Figure \ref{fig:apt-reconstruction-1}a of approximate stoichiometry V$_2$C.

As a result of partitioning, core austenite grains and necklace austenite grains will have different SFE. From the SEM-EDS measurements, the core austenite grains had a larger SFE than the necklace austenite grains due to the higher Al content \cite{Pierce2014,Saeed-Akbari2009}. M$_d$ temperature of the necklace austenite grains was lower and therefore more stable against transformation. However, the APT measurements showed that there was some variation in SFE between two necklace austenite grains, which may indicate a spread of SFE amongst necklace austenite grains. Nevertheless, the core austenite grains and necklace austenite grains still possess SFEs within the twinning regime \cite{Kim2016b,DeCooman2018}.

\subsection{Mechanical behaviour and deformation microstructure}

EBSD phase maps from interrupted tensile tests were obtained to investigate the microstructural evolution during tensile deformation in Figure \ref{fig:ebsd-vol-frac-tensile2}, and the engineering stress-strain tensile curve is shown in Figure \ref{fig:tensiles2}. The steel possessed a yield strength of 1090 $\usk\mega\pascal$, tensile strength of 1260 $\usk\mega\pascal$ and elongation of 54\%. It is evident from the true stress - true strain tensile curve (Figure \ref{fig:ebsd-vol-frac-tensile2}g) that the steel underwent two stages of hardening. In the first stage, Stage A ($\epsilon_t = 0.02 \, - \, 0.25$), the tensile curve was characterised by unsteady flow and a fluctuating strain hardening rate. In the second stage, Stage B ($\epsilon_t = 0.25 \, - \, 0.43$), there was steady hardening with a near constant hardening rate of 2.8 $\usk\giga\pascal$ up to the point of fracture.

Due to the difficulty in resolving ferrite and $\alpha$' martensite in EBSD, both ferrite and $\alpha$' martensite are coloured green in Figure \ref{fig:ebsd-vol-frac-tensile2}. However, $\alpha$' martensite can be distinguished from ferrite to some degree from correlative EBSD and SEM-EDS maps \cite{Field2018} as shown in Figure \ref{fig:sem-eds-mn-map}. Strain induced $\alpha$' martensite has a blocky morphology and possesses the same composition as the core austenite from which it transformed from as compared to ferrite which is equiaxed in morphology and is Mn depleted. Selected regions which could be confidently identified as strain induced $\alpha$' martensite are indicated by the white arrows in Figure \ref{fig:ebsd-vol-frac-tensile2}.

Unfortunately, due to the large strains involved, a significant area ($<$0.2) was not indexed in EBSD and so microstructural results will be interpreted and discussed qualitatively. From $\epsilon_t=0 \, - \, 0.28$, the austenite volume fraction steadily decreased while ferrite content steadily increased. This suggests that TRIP was occurring within the austenite phase and from Figure \ref{fig:ebsd-vol-frac-tensile2}b-c, the TRIP effect was observed to be mostly confined within the core austenite grains. At $\epsilon_t=0.43$, the indexing rate was very poor parallel to the transverse direction but improved drastically when scanned parallel to the normal direction. In Figure \ref{fig:ebsd-vol-frac-tensile2}f, it can be seen that the core austenite regions have completely transformed and any untransformed austenite was confined to within the necklace zones. When comparing the volume fractions obtained between Figures \ref{fig:ebsd-vol-frac-tensile2}d and f, it is evident that the ferrite content continued to increase, indicating continued TRIP. 

\begin{figure}[t!]
	\centering
	\includegraphics[width=\linewidth]{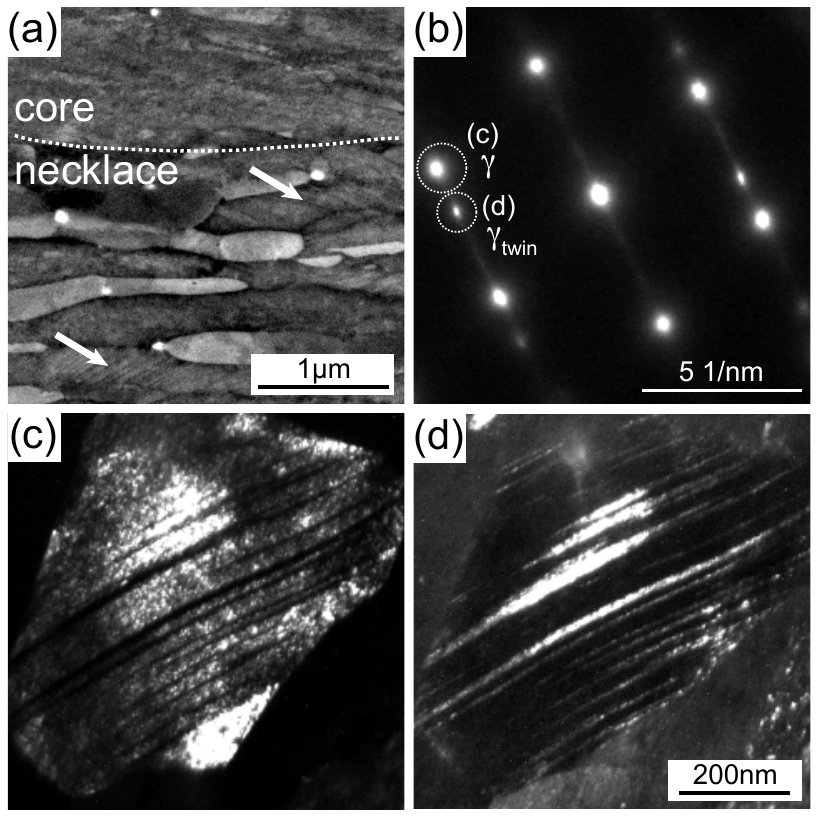}
	\caption{(a) BSE micrograph showing evidence of twinning in grains marked by the white arrow. (b) Diffraction pattern with a [110]$_\gamma$ beam direction showing matrix and twin spots obtained from a necklace austenite grain when deformed to $\epsilon_t=0.28$. Streaks were observed normal to the twin plane. Centre beam dark field micrograph obtained from the (c) austenite matrix spot and (d) twin reflection spot.}
	\label{fig:sem-tem-twins}
\end{figure}

In the specimens deformed to fracture and at $\epsilon_t=0.28$, twin-like features were observed in some of the necklace grains under BSE imaging, presumed to be necklace austenite. Twinning was indeed confirmed within the necklace austenite grains when a FIB foil was lifted out from the necklace region of a specimen deformed to $\epsilon_t=0.28$. Figure \ref{fig:sem-tem-twins} shows the evidence of twinning under BSE imaging and also confirmation of twins under TEM imaging.

\begin{figure}[t!]
	\centering
	\includegraphics[width=\linewidth]{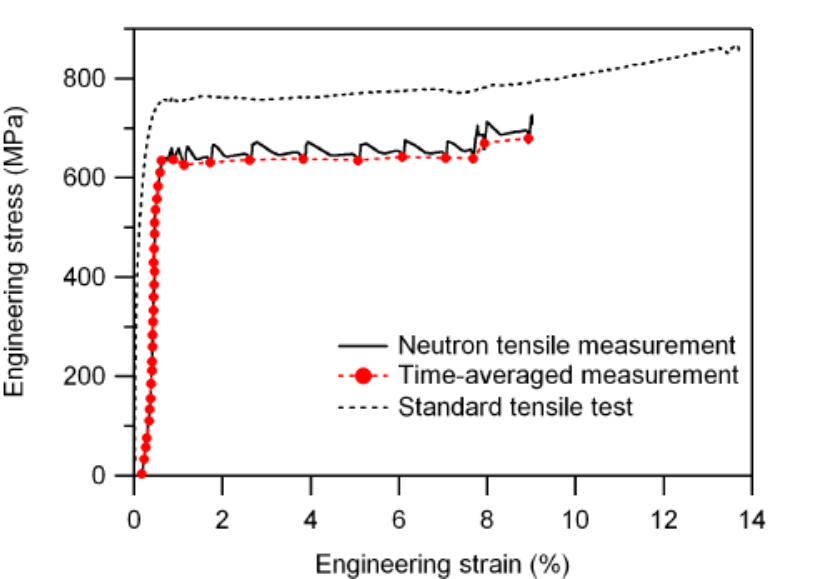}
	\caption{Engineering stress-strain curves of the tensile specimen used for in-situ neutron diffraction and of the sub-sized tensile specimen obtained from the same strip.}
	\label{fig:neutron-tensile}
\end{figure}

\subsection{In-situ neutron diffraction measurements}

Figure \ref{fig:neutron-tensile} shows the tensile curve obtained from the in-situ neutron diffraction tensile test, the time-averaged measurements and the tensile curve from the sub-sized specimen from the same ingot. It is noted that while the tensile properties obtained in the full sized and sub-sized neutron tensile specimens differed from those shown in Figure \ref{fig:tensiles2}, the two-stage strain hardening behaviour and the necklace-core microstructure was preserved (Figure \ref{fig:neutron-set-up-schematic}b). There are two likely reasons for the difference in tensile properties. The first reason was the difference in cooling rates after the homogenisation heat treatment. The ingot used to produce the tensile specimen for the in-situ neutron diffraction tensile test was furnace cooled, instead of water quenched, which allowed V to precipitate as carbides and coarsen significantly (Figure \ref{fig:neutron-set-up-schematic}b). The coarse precipitates would not be able to aid in grain refinement nor precipitation strengthening and possibly act as crack initiation sites. Secondly, warm rolling reductions were lower due to limitations of the laboratory rolling mill and the necklace layers were therefore not as developed. As a result, the neutron diffraction experiment was only able to proceed to approximately 9\% engineering strain which captured stage A type hardening and a small amount of stage B type hardening ($\epsilon=7.7$ to $8.9$\%).

\begin{figure}[t!]
	\centering
	\includegraphics[width=\linewidth]{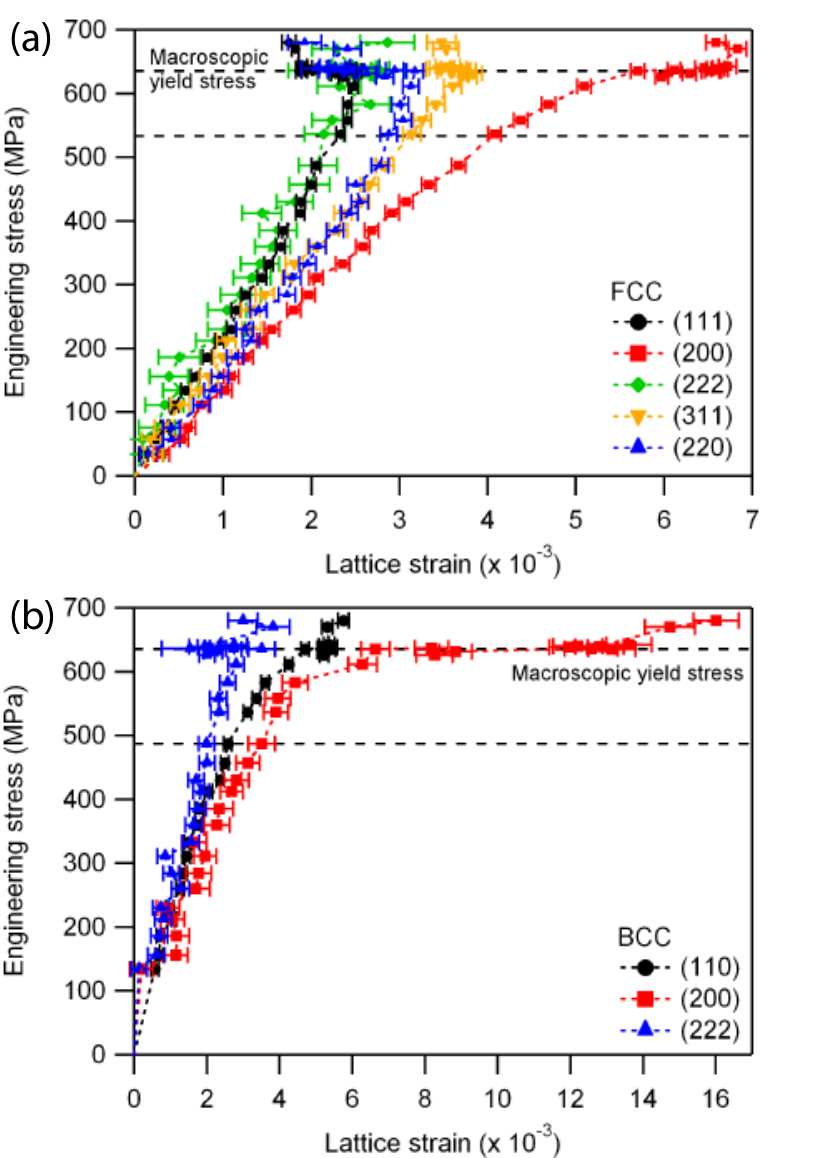}
	\caption{Lattice strain dependence on engineering stress of (a) FCC (b) BCC phases.}
	\label{fig:lattice-strain-neutron}
\end{figure}

Figure \ref{fig:lattice-strain-neutron} shows the stress dependence of lattice strain in the austenite (FCC) and ferrite (BCC) phases. The austenite peaks showed a linear increase in lattice strain with increasing stress up to approximately 430 MPa, whereupon the (220)$_\gamma$ peak showed a continuous increase in stress dependence with increasing strain up to the macroscopic yield stress. At 530MPa, the stress dependence of the (111)$_\gamma$ and (220)$_\gamma$ peaks decreased while the other peaks continued to increase linearly. At the macroscopic yield stress, all the FCC peaks experienced a decrease in lattice strain at constant stress before increasing again, with the exception of the (220)$_\gamma$ peak where the lattice strain continuously increased at the macroscopic yield stress and then decreased with increasing stress. 

\begin{figure}[t!]
	\centering
	\includegraphics[width=\linewidth]{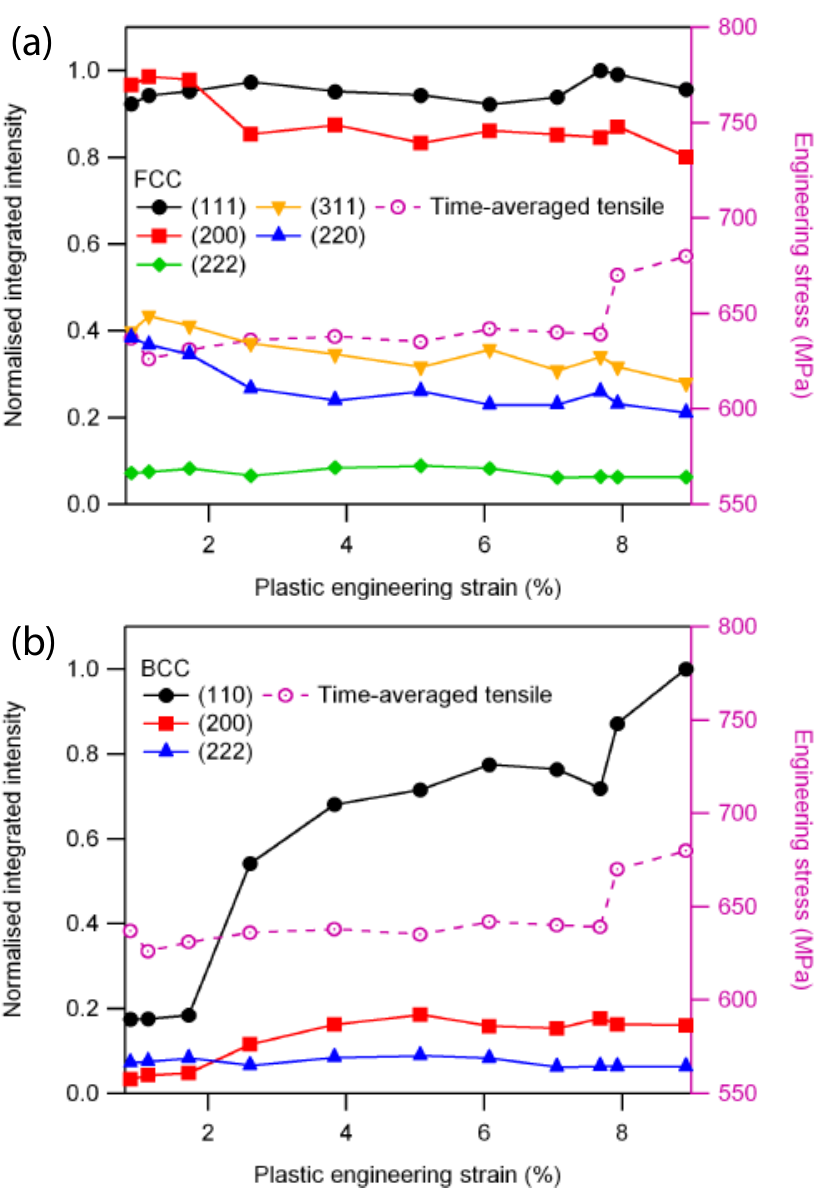}
	\caption{Normalised neutron diffraction integrated peak of (a) FCC and (b) BCC phases within the plastic strain region.}
	\label{fig:neutron-normalised-intensity-vs-strain}
\end{figure}

The BCC peaks showed a linear increase in lattice strain with increasing stress up to 490 MPa where the stress dependence of the (110)$_\alpha$ peak increased and at 590 MPa, the stress dependence of the (200)$_\alpha$ peak increased dramatically. At the macroscopic yield stress, the lattice strain of the (110)$_\alpha$ and (222)$_\alpha$ peak fluctuated at constant stress before increasing slightly at higher stresses. The (200)$_\alpha$ peak however, continued to increase in lattice strain dramatically at the macroscopic yield stress before increasing again with increasing stress at higher stresses. The ferrite produced by the martensitic TRIP will be of different composition and \textit{d}-spacing to the ferrite existing after IA, therefore limited reliance should be placed on the interpretation of the BCC lattice strains beyond the macroscopic yield point.

The elastic regions of the diffraction elastic constants (DECs) calculated from the curves in Figure \ref{fig:lattice-strain-neutron} are shown in Table \ref{tab:DEC}. It is evident that the austenite is stiffer than the ferrite phase overall, and therefore the austenite will shield the ferrite from stress to a certain extent.



\begin{figure*}[t!]
	\centering
	\includegraphics[width=\linewidth]{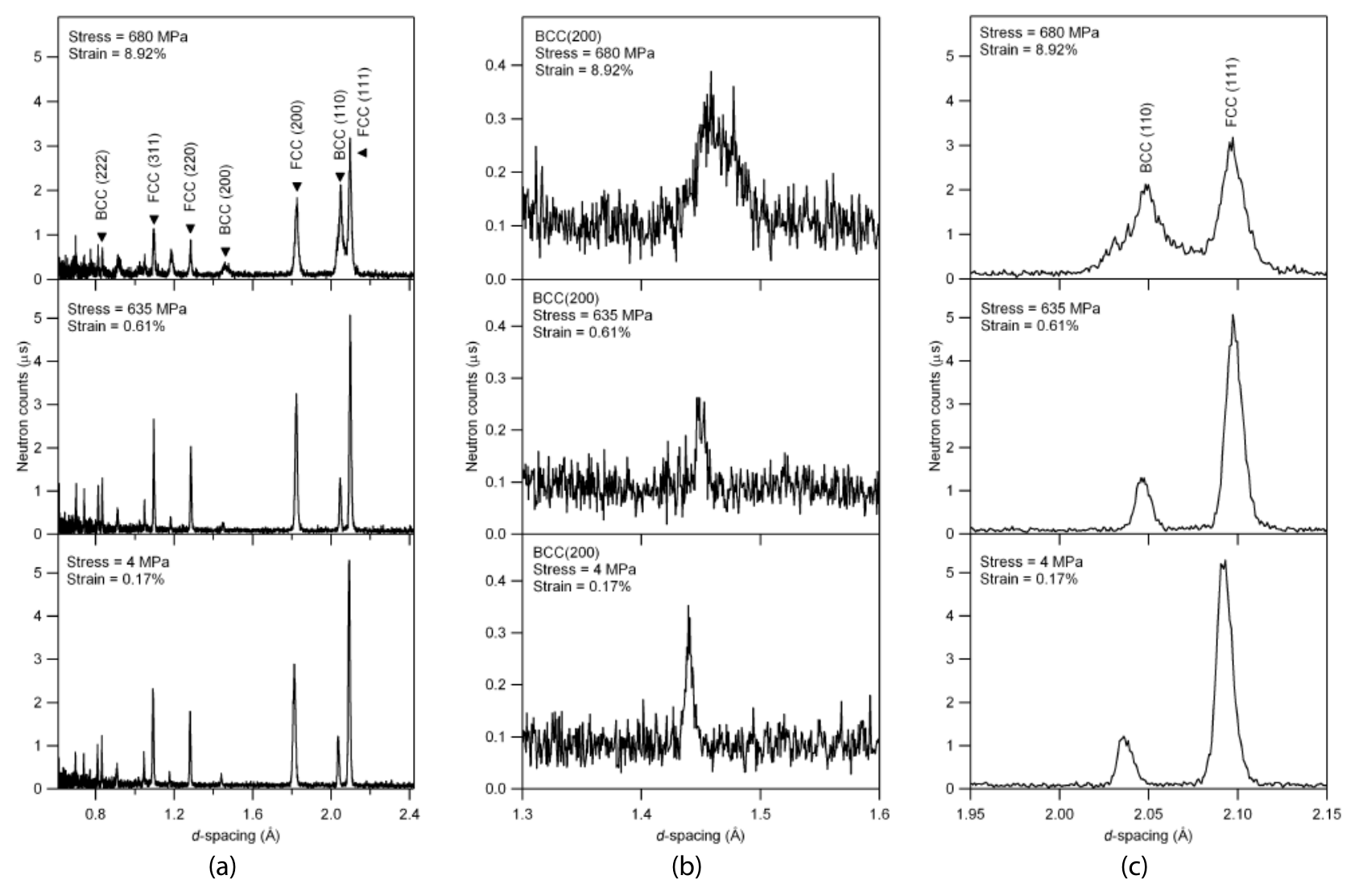}
	\caption{(a) Full neutron diffraction pattern and measured peaks when, from bottom to top, unstrained, at macroscopic yield point and fully deformed respectively. (b) BCC (200) peak. (c) BCC (110)$_\alpha$ and FCC (111)$_\gamma$ peak.}
	\label{fig:3-spectrums}
\end{figure*}

From Figure \ref{fig:neutron-normalised-intensity-vs-strain}, the integrated intensity of FCC peaks exhibited a slight decrease during initial plastic strain. While the maximum intensity has decreased (Figure \ref{fig:3-spectrums}c), the increase in width resulted in a near constant integrated intensity.  The BCC peaks exhibited more varied behaviour. The integrated intensity of the (200)$_\alpha$ peak increased slightly just before 2\% strain and increased gradually up to 5\% and remained fairly constant. Most notably, the (110)$_\alpha$ peak intensity increased significantly at 1.7\% strain and increased sharply again at 7.7\% strain, which corresponds to the beginning of stage B hardening. 

\begin{table}[h]
	\centering
	\caption{Table of DECs and anisotropy of resolved planes. The anisotropy of cubic crystals scale with 3$\Gamma$ where $3\Gamma = 3(h^2k^2+h^2l^2+k^2l^2)/(h^2+k^2+l^2)^2$ \cite{Daymond1997}. Uncertainties shown in parantheses.}
	\begin{tabular}{lcc}
		\toprule
		Plane & \multicolumn{1}{c}{Modulus (GPa)} &  \multicolumn{1}{c}{3$\Gamma$} \\
		\midrule
		(200)$_\gamma$ & 139 (3)   & 0 \\
		(311)$_\gamma$ & 164 (3)   & 0.47 \\
		(220)$_\gamma$ & 176 (4)   & 0.75 \\
		(111)$_\gamma$ & 224 (4)   & 1 \\
		(222)$_\gamma$ & 216 (9)   & 1 \\
		\midrule
		(200)$_\alpha$ & 119 (9)   & 0 \\
		(110)$_\alpha$ & 169 (7)   & 0.75 \\
		(222)$_\alpha$ & 182 (16)   & 1 \\
		\bottomrule
	\end{tabular}%
	\label{tab:DEC}%
\end{table}%

\section{Discussion}

\subsection{Processing and microstructure}
The investigated necklace-core type microstructure arose from a continuous hot roll - warm roll -  anneal process. Necklacing or partial recrystallisation is common in the hot or warm deformation of stainless steel \cite{Dehghan-Manshadi2008} but has not been commonly reported in medium Mn steel. He \textit{et. al.} \cite{He2018} and Hu \textit{et. al.} \cite{Hu2019} obtained a similar partially recrystallised necklace-core type microstructure in a Fe-10Mn-2Al-0.5C-0.7V steel that was hot rolled, cooled to room temperature and subsequently warm rolled at various temperatures and reductions followed by a long IA of 5 h. While the mechanical properties were different, the resulting strain hardening behaviour was remarkably similar with stage A and B-type strain hardening, suggesting the importance of microstructure on the tensile behaviour of medium Mn steels with necklace-core type microstructures. 

The key processing step was the warm rolling step at 700 $\degree$C where several processes were observed to occur: nucleation of ferrite and recrystallisation of austenite, \textit{i.e.} formation of the necklace and element partitioning. It is known that ferrite transformation from austenite is sluggish when austenite is held isothermally within the intercritical temperature regime \cite{Nakada2014}. Warm rolling introduces defects as nucleation sites for ferrite formation, which explains why ferrite was largely observed to be located in the necklace layer (\textit{i.e.} PAGB) and along shear bands within the prior austenite grain. Because austenite grains were also found in the necklace layer, recrystallisation of austenite was thought to be occuring simultaneously with ferrite transformation \textit{via} the necklace recrystallisation mechanism. As warm rolling progressed, more necklace layers were added, consistent with the findings by Ponge and Gottstein \cite{Ponge1998}. However, recrystallisation was incomplete, resulting in austenite cores which are likely highly deformed. The core austenite grains were likely able to recover during the IA at 700 \degree C but are still presumed to have a much higher dislocation density compared to the necklace grains, contributing to the high yield strength of the steel. 

Partitioning was also thought to be enhanced by warm rolling. From Figure \ref{fig:sem-eds-mn-map} and Table \ref{tab:SEM-EDS-ICP-compositions}, the necklace austenite grains were observed to have a significantly higher Mn content than the core austenite grains. APT measurements also showed that the Mn content in the necklace austenite grains were very close to the thermodynamic equilibrium. A possible reason was due to the small grain sizes of the necklace grains which considerably shortens the diffusion distance between neighbouring ferrite grains \cite{Lee2011b}. Another reason could be due to pipe diffusion of Mn \cite{KwiatkowskidaSilva2017,LIN2016}. As the necklace grains formed as warm rolling progressed, it is likely that these grains also experienced a certain degree of deformation.  While not enough to trigger recrystallisation within the necklace, the dislocation density is held to be sufficient to enhance partitioning between austenite and ferrite within the necklace via pipe diffusion. Necklace grains which formed early during warm rolling would have experienced a larger amount of warm deformation as compared to the necklace grains which formed later on or during the final IA. As a consequence, it is likely that partitioning would have occurred to a greater extent in the early necklace austenite grains. This effect may explain the slight difference in compositions between necklace austenite grains measured by APT shown in Table \ref{tab:SEM-EDS-ICP-compositions} and Figure \ref{fig:apt-reconstruction-1}.

\subsection{Tensile behaviour and microstructure}
The two-stage tensile behaviour of the steel was investigated through an in-situ neutron diffraction tensile test and the evolution of microstructure during tensile deformation was investigated with EBSD through a series of interrupted tensile tests. 

Stage A of the strain hardening curve is characterised by negligible increase in engineering stress with applied strain and a serrated strain hardening rate. Many researchers associate this post-yielding serration with the TRIP effect \cite{Han2014,Zhang2017b,Hu2019}. Zhang \textit{et. al.} \cite{Zhang2017b} demonstrated that before uniform elongation, retained austenite undergoes the TRIP effect caused by the passage of a L\"{u}ders band. The increase in BCC (110)$_\alpha$ peak intensity in Figure \ref{fig:neutron-normalised-intensity-vs-strain}b also indicates that martensitic transformation was active.

Microstructurally, it can be seen from Figure \ref{fig:ebsd-vol-frac-tensile2} that the TRIP effect in stage A was largely confined within the core austenite grains. This can be explained by the difference in austenite stability between the core and necklace austenite grains (SEM-EDS measurement in Table \ref{tab:SEM-EDS-ICP-compositions}). The core austenite having less Mn and a significantly larger grain size is less stable and more likely to TRIP. The core austenite also possess a grain size of $>$10 $\mu$m, and will therefore be less strong than the necklace austenite which, like the necklace ferrite, is much smaller than 1 $\mu$m in size. It should be noted that M$_d$ calculations from Table \ref{tab:SEM-EDS-ICP-compositions} were based on empirical calculations based on austenitic stainless steels dating back to Angel \cite{Angel1954} in 1954 and Nohara \textit{et. al.} \cite{Nohara1977}  in 1977 and may not be fully applicable in medium Mn steels. M$_d$ temperatures in Table \ref{tab:SEM-EDS-ICP-compositions} should therefore be interpreted conceptually.

It is also noteworthy that while the SFE of the austenite core was within the twinning region, TRIP rather than TWIP was observed. Several factors could be at work. The SFE calculation may be erroneous as the empirical model \cite{Pierce2014} used in this study was not developed for use in medium Mn steel but more importantly, the deformation mode in medium Mn steels cannot be determined by SFE alone, unlike TWIP steels which do not transform \cite{Galindo-Nava2017}. In this scenario, the austenite stability of the core austenite grains was such that strain induced martensitic transformation was preferable to twinning. 

During stage A hardening, the FCC peaks in Figure \ref{fig:lattice-strain-neutron}a, with the exception of (220)$_\gamma$ experienced a decrease in lattice strain, indicating that a stress-relieving mechanism, \textit{i.e.} TRIP, was active. This result was also observed by Zhang \textit{et. al.} \cite{Zhang2017b} in a medium Mn TRIP steel. It is likely that load partitioning after yielding took place with the martensite which formed from the core austenite grains taking on the majority of the load. It is worth noting from Figure \ref{fig:neutron-normalised-intensity-vs-strain}b that the intensity of BCC (110)$_\alpha$ peak increased rapidly at 1.7\% strain but slowed at 7.7\% strain which likely indicates the decreasing transformation rate of the core austenite grains.

Stage B of the strain hardening curve is characterised by a more uniform but slightly fluctuating strain hardening rate. The core austenite having transformed to core martensite now strengthens the steel which may explain why the strain hardening rate in stage B is on average higher than stage A. It is likely that in stage B, the load is now partitioned from the core martensite grains to the necklace grains which now begin to deform. The necklace austenite grains are of interest as both TWIP and TRIP behaviour were observed in stage B. The twins observed in a necklace austenite grain in Figure \ref{fig:sem-tem-twins} support the suggestion that TWIP was active while the increase in ferrite/$\alpha$' martensite fraction in Figure \ref{fig:ebsd-vol-frac-tensile2}h and the second increase in intensity of the BCC (110)$_\alpha$ peak at 7.7\% strain in Figure \ref{fig:neutron-normalised-intensity-vs-strain}b support the suggestion that TRIP was active. It is therefore reasonable to conclude that both TWIP and TRIP mechanisms were active in the necklace austenite grains during stage B. However it was not determined if the successive TWIP+TRIP effect was active. 

Several reasons may be possible for the difference in deformation mechanisms in the necklace austenite grains. From Table \ref{tab:SEM-EDS-ICP-compositions}, there is a spread in composition in the necklace austenite grains resulting in dfferences in SFE and austenite stability. It has also been reported that grain size could be a determining factor whether an austenite grain deforms by TWIP or TRIP in medium Mn steels \cite{Lee2015d,Sohn2014a}, however other factors such as orientation of austenite grains relative to the loading direction \cite{Xu2017} may affect which deformation mechanism activates in any given grain. As the necklace austenite composition was not at the thermodynamic equilibrium as seen in Table \ref{tab:SEM-EDS-ICP-compositions}, differences in composition between necklace austenite grains may also account for variations in SFE and austenite stability and therefore influence the type of active deformation mechanism.

\subsection{Industrial relevance}
Researchers and automakers are aiming to develop new steel grades with $>$1000 $\usk\mega\pascal$ tensile strength and 20\% elongation termed as the 3\textsuperscript{rd} generation Advanced High Strength Steels (AHSS) \cite{Rana2016}. WorldAutoSteel \cite{Design2011} postulates a TWIP550/900 grade ($500-550 \usk\mega\pascal$ yield strength, $980-990 \usk\mega\pascal$ tensile strength and $50-60$\% elongation) in the design of the Future Steel Vehicle (FSV). Many medium Mn steels in the literature exceed this target, demonstrating the potential of this family of steels. However, medium Mn steels will still face industrial concerns such as cost, formability and weldability \cite{Rana2016}. 

\begin{figure}[t!]
	\centering
	\includegraphics[width=\linewidth]{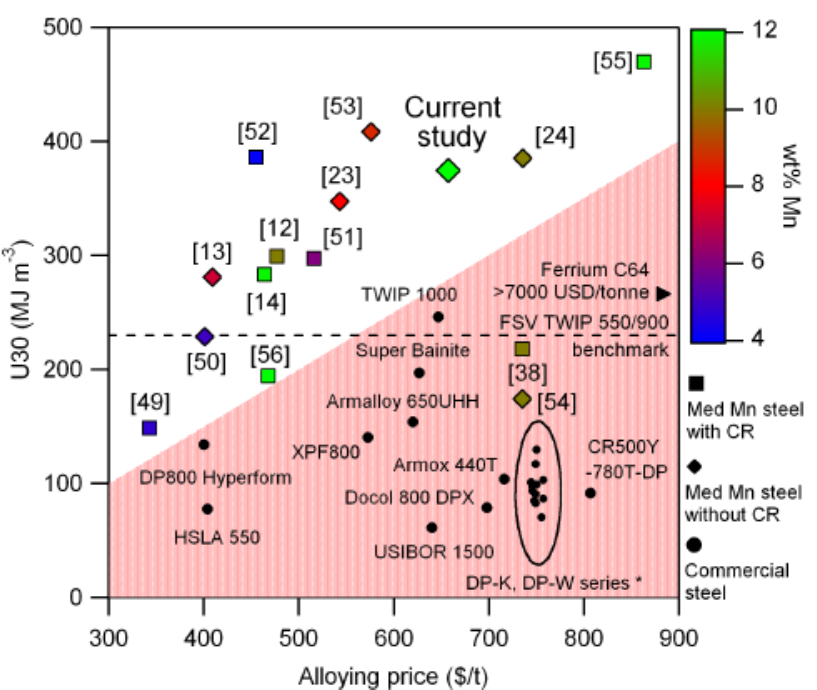}
	\caption{Comparison of cost and energy absorbtion (U30) between commercial steels and medium Mn steels in the literature. CR - Cold Rolling. *ThyssenKrupp dual phase steel product catalogue.}
	\label{fig:u30}
\end{figure}

At this stage of alloy development, we introduce a comparator for performance, U30, which represents the energy absorbed during tensile deformation which is determined by the area under the tensile curve up to $\epsilon_t=0.3$, where 0.3 is determined to be the limit of useful ductility. As medium Mn steels are designed for energy absorbing applications, U30 would serve as a useful estimate. In Figure \ref{fig:u30}, several commercial steels used in the automotive and armour industries are compared to medium Mn steels in the literature based on price and U30. Steel prices were determined through prices of elements, obtained from www.metalbulletin.com and The London Metal Exchange. Prices are correct as of May 2019 and calculated on the basis of steel scrap with the addition of ferroalloys or pure elements wherever appropriate. Cost of production was not factored but is acknowledged to be a significant factor in the overall price. Medium Mn steels are subdivided depending on whether cold rolling was utilised in their processing. The issue with cold rolling followed by heat treatment is that this necessitates a batch production process which is expensive to realise commercially.

It can be seen that most commercial steels lie within the shaded region and most medium Mn steels lie above, which indicates that most medium Mn steels have better energy absorbtion capabilities for the same alloying price. Many medium Mn steels also have U30 values greater than the FSV TWIP550/900 benchmark, which demonstrates the readiness of medium Mn steels in terms of mechanical properties. In the current study, U30 was relatively high but also in price. In order to lower the price, the use of expensive elements such as V needs to be reduced. Mn content also needs to be lowered to $\leq 8$ wt\% for cost and processability reasons. The aim of our future work will be to reduce the price of the alloy to  $400-500$ \$/t, while keeping the same U30 and a simple thermomechanical process.

\nocite{Luo2015}
\nocite{Shao2017}
\nocite{Hu2017b}
\nocite{Lee2016}
\nocite{Lee2014}
\nocite{DeCooman2017}
\nocite{Lee2015c}
\nocite{Lee2017}
\nocite{Zhu2017}
\nocite{Hu2019}
\nocite{He2018}
\nocite{He2018a}
\nocite{Sohn2017}
\nocite{Benzing2019}

\section{Conclusions}
In conclusion, a 12 wt\% Mn medium Mn steel with an engineering yield stress of 1.09 GPa and true ultimate tensile strength of 1.9 GPa was manufactured and processed in a continuous method which consisted of hot and warm rolling followed by a 30 min intercritical anneal. The steel possessed a partially recrystallised or necklace-core microstructure with coarse austenite grains forming the core and fine-grained ferrite and austenite grains forming the necklace. It was found using EBSD, TEM and in-situ neutron diffraction that the steel had a two-stage hierarchical deformation behaviour where the core austenite grains first transform to martensite and the necklace austenite grains subsequently deform by transformation and twinning-induced plasticity. Its deformation behaviour resulted in a ductility of 54\% and sustained hardening throughout the tensile curve, with an energy absorbtion capability per unit alloy cost competitive with the best model steels in the literature.

\section{Acknowledgements}
The authors would like to acknowledge the support from A*STAR, the Royal Society, UK Ministry of Defence and EPSRC grant number EP/L025213/1. We would also like to thank Felicity Dear and Felix Wolfgang Liwerski for their contributions.

\section{References}
\bibliographystyle{Acta}
\bibliography{library.bib}

\end{document}